\documentstyle[aaspp4]{article}
\begin{document}
\title{\bf Possible Detection of Baryonic Fluctuations in the Large-Scale
Structure Power Spectrum}

\author{Christopher J. Miller and Robert C. Nichol}
\affil{Department of Physics, Carnegie Mellon University, 5000 Forbes Ave.,
Pittsburgh, PA-15232 (chrism@cmu.edu \& nichol@cmu.edu)} 

\author{David J. Batuski} 
\affil{Department of Physics \& Astronomy, The University of Maine, Orono,
Maine, ME-04469 (david.batuski@umit.maine.edu)}

\begin{abstract} 
We present a joint analysis of the power spectra of
density fluctuations from three independent cosmological redshift surveys; the
PSCz galaxy catalog, the APM galaxy cluster catalog and the
Abell/ACO cluster catalog. Over the range $0.03 \le k \le 0.15h$Mpc$^{-1}$,
the amplitudes of these three power spectra are related through a simple
linear biasing model with $b = 1.5$ and $b = 3.6$ for Abell/ACO versus APM and
Abell/ACO versus the PSCz respectively. Furthermore, the shape of these power
spectra are remarkably similar despite the fact that they are comprised of
significantly different objects (individual galaxies through to rich
clusters).  Individually, each of these surveys show visible
evidence for ``valleys'' in their power spectra -- {\it i.e.}  departures from
a smooth featureless spectrum -- at similar wavenumbers. 
We use a newly developed statistical technique called the
{\it False Discovery Rate}, to show that these ``valleys'' are statistically
significant. One favored cosmological explanation for such features in the
power spectrum is the presence of a non-negligible baryon fraction ($\Omega_b$)
in the Universe which causes acoustic oscillations in the transfer function of
adiabatic inflationary models.  We have
performed a maximum-likelihood marginalization over four important cosmological
parameters of this model ($\Omega_m$, $\Omega_b$, $n_s$, $H_o$). We
use a prior on $H_0 = 69\pm{15}$, and find
$\Omega_mh^2 = 0.12^{+0.03}_{-0.02}$, $\Omega_bh^2 =0.029^{+0.011}_{-0.015}$,
$n_s = 1.08^{+0.17}_{-0.20}$ (2$\sigma$
confidence limits) which are fully consistent with the favored values of these
cosmological parameters from the recent Cosmic Microwave Background (CMB)
experiments. This agreement strongly suggests that we have detected baryonic
oscillations in the power spectrum of matter at a level expected from a Cold
Dark Matter (CDM) model normalized to fit these CMB measurements.

\end{abstract}


\section{Introduction}
We present a new analysis on the power spectra of
density fluctuations ($P(k)$) as derived from three recently available
independent cosmological redshift surveys; the Abell/ACO Cluster Survey as
defined in Miller \& Batuski (2001) and Miller et al. (2001a), the IRAS
Point Source redshift catalog (PSCz; Saunders et al. 2000), and the Automated
Plate Machine (APM) cluster catalog (Dalton et al. 1994).  For the first time,
the volumes traced by these surveys are large enough to accurately probe the
power spectrum to wavenumbers of $k = 0.015$ (Abell/ACO), $0.025$ (PSCz) and
$0.030h$Mpc$^{-1}$ (APM). Throughout, we use $h=H_o/100$km s$^{-1}$Mpc$^{-1}$.

Such information on the large--scale distribution of matter in the Universe is
critical for constraining cosmological models of structure formation as well
as determining the cosmological parameters.  For example, the amplitude and
shape of $P(k)$ below $k \sim 0.05h$Mpc$^{-1}$ can be used to discriminate
between a high and low value of $\Omega_mh$, while a non-negligible baryon
fraction ($\Omega_b$) would produce noticeable oscillations in $P(k)$ at $k <
0.1h$Mpc$^{-1}$ (with the oscillations becoming broader, and more easily
detectable, toward smaller $k$; see Eisenstein et al. 1998). Cosmological
constraints based on the large--scale structure (LSS) in the Universe are
independent and complementary to those derived from the Type Ia supernovae and
Cosmic Microwave Background (CMB) experiments (see Bond et al. 2000; Jaffe et
al. 2000) and thus breaking key degeneracies inherent in these other
cosmological measurements (see, for example, Tegmark, Zaldarriaga \& Hamilton
2001).

In the past, cosmological studies of the power spectrum of density
fluctuations have been hampered in three ways; {\it i)} uncertainties in the
shape of the $P(k)$ on very large scales, {\it ii)} the form of the relative
biasing between the luminous and dark matter, and {\it iii)} the possible
existence of a narrow ``bump'' in the $P(k)$ (Landy et al. 1996; Einasto et
al. 1997). As we will show in Sections \ref{bias} and \ref{fluc}, our new
data-sets allow us to address these concerns and thus facilitate a more robust
determination of the cosmological parameters from LSS measurements. Our work
differs from other recent attempts to constrain cosmological models using LSS
data (Novosyadlyj et al. 2000; Tegmark et al. 2001; Efstathiou \& Moody 2000;
Huterer, Knox \& Nichol 2000) since we first focus on the detection and
interpretation of features in the matter power spectrum, followed by parameter
estimation based on the favored models that explain these features.

\section{Biasing}
\label{bias}
In Figure 1 (top) we plot the $P(k)$ for our three samples. The Abell/ACO
power spectrum is from Miller \& Batuski (2001), the APM result is from
Tadros, Efstathiou, and Dalton (1998), and the PSCz data are from Hamilton \&
Tegmark (2000).  In all three plots, we exclude any data with errors $> 50$\%
of the power for obvious reasons but we note that their inclusion makes no
differences in our final results. The errors are all $1\sigma$ as quoted by
the authors.  In the bottom frame, we plot the same data for the three samples
after shifting the amplitudes of the APM and PSCz surveys to match that of the
Abell/ACO survey. We have applied various techniques to calculate this
amplitude shift ({\it e.g.} a $\chi^2$ minimization of the data with nearly
identical $k$-values as well as using model fits to the data and
re--normalizing them to the Abell/ACO data), but in all cases we obtain nearly
identical results: the relative bias between the three samples over the range
$0.03 \le k \le 0.15h$Mpc$^{-1}$ is $b = 1.5$ and $b = 3.6$ for Abell/ACO
versus APM and Abell/ACO versus the PSCz respectively.

A remarkable aspect of Figure 1 is the overall success of a simple linear
biasing model in re-normalizing the amplitudes of these power spectra over
nearly a decade of scale. A scale--independent biasing model, over the scales
discussed herein, has already been predicted from recent numerical simulations
(Narayanan, Berlind \& Weinberg 2000) and therefore, allows us to confidently
re-scale these three different power spectra, thus facilitating the detection
of features in the combined $P(k)$ as discussed below.

\section{The Shape of the Power Spectrum}
\label{fluc}

The overall shape of our combined power spectrum is shown in Figure 1 and it
is unique for two reasons. First, we see no evidence at small $k$ values for a
turn--over in the $P(k)$ toward a scale-invariant spectrum as previously
hinted at in other LSS analyses (Tadros \& Efstathiou 1996; Peacock 1997;
Gatza\~{n}aga \& Baugh 1998). This large-scale power has also been
witnessed in other recently reported $P(k)$ measurements (see Efstathiou \&
Moody 2000; Schuecker et al. 2001).  Such large--scale power in the $P(k)$
indicates a low value for the shape parameter ($\Gamma = \Omega_m\,h < 0.3$)
since this has the effect of sliding the matter power spectrum to the smaller
$k$ values compared to a critical matter density universe.  Secondly, we do
not find a narrow ``bump'' in the $P(k)$ as reported by Einasto et al. (1997)
and Landy et al. (1996) but instead witness ``valleys'' in the power spectrum.
However, these previous surveys did not have the volume to see the large scale
($k<0.03$) power in $P(k)$ and thus only saw the down--turn of the ``valley''
giving the appearance of a ``bump'' in the power spectrum at $k\simeq0.05$ and
therefore, our $P(k)$ may still be consistent with the Landy et al. and the
Einasto et al. power spectra.  For the remainder of this section, we focus on
the the two ``valleys'' we see in Figure 1 at $k\simeq 0.035 $ and $k\simeq
0.090h$Mpc$^{-1}$.

\subsection{False Discovery Rate} 

In this section, we investigate the statistical significance of the two
features seen in Figures 1 and 2. We wish to determine if all of the data points are
consistent with being drawn from a smooth, featureless, power spectrum.  In
the statistical literature, this is known as multiple hypothesis testing since
one is testing, for each point, the null hypothesis that it was drawn from a
featureless $P(k)$. The key issue then becomes choosing the threshold (in
probability) which these null hypotheses are rejected.

Traditionally, this is done by rejecting all points that are above a certain
$\sigma_{rej}$ threshold.  Unfortunately, there is a major problem with this
methodology since the number of data points that are mistakenly rejected
depends on the size of the data-set. For example, if all our 37 data points
were uncorrelated and truly drawn from a smooth $P(k)$,
we would expect, on average, only 1.75
of these data points to be rejected (and thus in error) for a $\sigma_{rej}=2$
threshold.  However, if we had one million data points, then the number
mistakenly rejected data points at the $\sigma_{rej}=2$ level would be
approximately 50,000. [Note: this comparison only works if all of the real
data were uncorrelated.]  To guard against the over-detection of
false discoveries, one could arbitrarily
increase the significance threshold to $\sigma_{rej}=4$, thus reducing the
number of errors but would lead to a much more conservative test. This is not
to say that all of the $2\sigma$ rejections would be wrong, but simply that
you would have many more false rejections. Thus, any significance threshold is
arbitrary and highly dependent on the data size. So, enforcing a
$\sigma_{rej}=3$ threshold for small data-sets can be overly conservative. In
summary, the more tests one does, the more stringent the required threshold becomes
to avoid making too many false discoveries.

Ideally, we need a statistical technique that is more adaptive and whose
interpretation does not depend on the data size. Instead of $\sigma_{rej}$,
we will choose  to
control the false
discovery rate
($\alpha$)-- which is defined to be the percentage of mistakenly rejected points
out of the total number of points rejected.  This is clearly independent of
the data size. Such an adaptive statistical tool is the {\it False Discovery
Rate} (FDR; Benjamini \& Hochberg 1995).
Once we choose $\alpha$, then FDR defines an appropriate significance
threshold to obtain this false discovery rate for the dataset in question.  For example, if
we choose $\alpha = \frac{1}{4}$ and reject eight data points,
then on average, only two of these points are in error even though their
significance (as implied by their $\sigma$'s) may appear low. Again,
arbitrarily setting $\sigma_{rej}=3$ for our data-set may be too conservative.
Instead, by {\it a priori} controlling the false discovery rate,
we can state with
statistical confidence that six out of eight rejected data points are true
outliers against the null hypothesis. We briefly discuss {\it FDR} here and
refer the reader to 
Nichol et al. 2000 and Miller et al. 2001b for further details.
  
Operationally, we first compute the $p$--value$\footnote{The
p--value is the probability that sampling from an ensemble of datasets
would lead to a data value with an equal or higher deviation from the
the null hypothesis.}$ for each test. Herein we have used a
a smooth CDM model based on our best-fit cosmological parameters
(Section \ref{fits}) but with the baryon oscillations removed (see Eisenstein
\& Hu 1998). However, the reader should note that this model is nearly
an exact power-law over the range $k > 0.02h$Mpc$^{-1}$, and so
our results would not change if we used a simple power-law fit
to the data (without the valleys). 
We then rank, in increasing size, the $p$-values (for each test)
and draw a line of slope $\alpha$ and zero intercept.
Recall, $\alpha$ is the maximum acceptable false discovery rate. The
first crossing of this line with a $p$--value (moving from larger to smaller
$p$--values) defines the significance threshold $\sigma_{rej}$, below which
all points are rejected based on our null hypotheses. On average, only
$\alpha\times100\%$ of these rejected points will be in error.

Figure 2 presents the result of applying FDR to our combined $P(k)$ using
an uncorrelated dataset (as opposed to the correlated data in Figure 1).
Specifically, we use the uncorrelated $P(k)$ given
by Tegmark \& Hamilton (2000) for the PSCz while Tadros et al. (1998) claim
their APM $P(k)$ is uncorrelated and thus we use their data points
directly. For the Abell/ACO catalog, Miller \& Batuski (2001) have
shown that their $P(k)$ is uncorrelated for separations of $\Delta k \simeq
0.015h$Mpc$^{-1}$. Therefore, as can be seen from Figure 1, the data at
$k>0.04$ is already uncorrelated while for smaller $k$ values we simply
re--sample the data in such a way that the minimum separation between points
is at least $\Delta k \sim 0.01h$Mpc$^{-1}$.
In Figure 2, the circled points are rejected (based on our null hypotheses) with a false
discovery rate
of $\alpha = 0.25$, while the points outlined with squares are rejected with a
$\alpha=0.10$. 

We detect the ``valleys'' at both $k \sim 0.035h$Mpc$^{-1}$ and
at $k \sim 0.09h$Mpc$^{-1}$.  The power of the false discovery rate is
that it ensures that no more than $25\%$ of the eight
rejections (circles) could be incorrect. If we apply the much more stringent
constraint of $\alpha = 0.1$, we only reject only one point
(at $k \sim 0.09h$Mpc$^{-1}$), but FDR limits the number of false
rejections in this case essentially  to zero.
This allows us to state with statistical
confidence that the fluctuations are true outliers against a smooth,
featureless spectrum. Note that each of the three data sets contributes
to the features, and so the detection is not dominated by one sample.
In the next two sections, we review possible
explanations for these observed features in our $P(k)$ including systematic
uncertainties and cosmological effects.

\subsection{Systematic Uncertainties}
 
In this section, we consider both measurement error and sampling effects as
possible explanations for the features seen in the power spectra shown in
Figures 1 \& 2.  To address the first of these, we simply note that Miller \&
Batuski (2001) used several different methods of calculating the $P(k)$ for
the Abell/ACO catalog and observed no significant differences in
the measured $P(k)$ for $k \le 0.02h$Mpc$^{-1}$.  Further evidence that the
fluctuations seen in Figures 1 \& 2 are not the result of the methodology
comes from the fact that the authors of the three power spectra all used different
methodologies to calculate their $P(k)$; the APM survey was analyzed using the method
of Feldman et al. (1994), the analysis of the Abell/ACO catalog
followed Vogeley et al. (1992) and Feldman et al. (1994), while the PSCz
$P(k)$ was derived from a Karhunen - Loeve (KL) eigenmode analysis.

Next, we consider sampling effects {\it e.g.}  could artifacts of the design
and construction of these surveys have produced such features and are the
surveys independent and representative of the whole Universe?  We believe such
effects are highly unlikely for two reasons. First, each of these three
surveys was constructed in a different way and thus possess significantly
different window functions. For example, the APM survey only covers a
steradian of the sky centered on the Southern Galactic Cap, with a near
constant number density of clusters/groups over the redshift range $15000 \le
cz \le 35000$km s$^{-1}$, while the Abell/ACO cluster sample covers over
2$\pi$ steradian with a near constant number density of rich clusters
out to $cz= 42000$km s$^{-1}$ (in the north) and $cz= 30000$km s$^{-1}$ (in
the south). These two cluster surveys are independent of each other since
$\simeq90\%$ of the APM clusters used by Tadros et al. (1998) are non--Abell
systems and are thus not in the Abell/ACO sample (Miller \& Batuski 2001). In
contrast, the PSCz galaxy redshift survey covers 10.6 steradian (84\% of the
entire sky) and has a number density that falls off steeply beyond
$cz=12000$km s$^{-1}$. Therefore, these three surveys sample different volumes of the
Universe, use different tracers of the mass (from galaxies through to rich
clusters) and are independent of each other.

We stress here that these features are seen in all of the individual power
spectra at similar wavenumbers and therefore, not a artifact of combining the
three $P(k)$'s, which we did simply to increase the overall statistical
significance of these ``valleys''. This concordance is a powerful consistency
check which argues against statistical and systematic uncertainties producing
these features. Moreover, the volume sizes of these three surveys are so large
that we hope to have reached a ``fair sample'' of the Universe and thus these
features can not be explained away as unusual and only present in our parts of
the Universe.  We therefore believe that these fluctuations in the $P(k)$ are
physical and in the next section we review possible cosmological explanations
for them.

\subsection{Cosmological Explanations for the Fluctuations}

One possible cosmological explanation for these ``valleys'' in the observed
$P(k)$ is the existence of corresponding features in the initial power
spectrum of density fluctuations coming out of Inflation. This explanation has
been proposed for the excess power or correlations seen in several of previous
surveys (Broadhurst et al. 1990; Landy et al. 1996; Einasto et
al. 1997). Unfortunately, the physical mechanism for producing such features
in the initial power spectrum remains unclear (see Atrio-Barandela 2000;
Einasto 2000).
  
A more natural and well--understood explanation is the presence of a
non-negligible baryon fraction in the Universe which leads to a coupling (at
redshifts $z \ge 1000$) between the CMB photons and the baryonic matter thus
resulting in acoustic oscillations which leave an imprint on the matter power
spectrum (see Eisenstein \& Hu 1998 and references therein).  
Recently, Eisenstein et al. (1998) examined this cosmological model and tested
it against three LSS data sets; the APM de--projected $P(k)$ of Gazta\~{n}aga \&
Baugh (1998), the $P(k)$ compilation of Peacock \& Dodds (1994), and $P(k)$
from the Abell/ACO sample from Einasto et al. (1997). Only the Einasto et
al. $P(k)$ had a noticeable feature (``bump'') in the power spectrum and
Eisenstein et al. (1998) were unable to find a satisfactory cosmological model
that fitted these data. Their analysis indicated two equally likely fits to
the data, one with $\Omega_m < 0.2$ and the other with $\Omega_m > 0.7$, while
both models needed $\Omega_b/\Omega_m \simeq 0.3$. The high $\Omega_m$ model
was excluded by the Big Bang Nucleosynthesis upper limit of $\Omega_bh^2 <
0.026$ (Burles, Nollett, and Turner 2000) while the low $\Omega_m$ model was
rejected by the lack of large--scale power seen in the $P(k)$ for $k <
0.05h$Mpc$^{-1}$.

We present here new constraints on model $P(k)$ using much improved datasets
than those used by
Eisenstein et al. (1998). The improvement in the data comes from the
larger volumes traced by these surveys, thus allowing smaller $k$'s to
be probed with a higher resolution.  In the next section, we re-examine this
scenario and find that baryonic oscillations match with the power spectra in
Figure 1.  We note here that Tegmark et al. (2001) also hinted at the possible
detection of baryon fluctuations in the PSCz $P(k)$ but a statistical analysis
was not performed.

\section{Cosmological Parameter Estimation}
\label{fits}

We have used the cosmological models of Eisenstein \& Hu (1998) to perform a
parameter estimation which can be compared with the recent CMB results
(Tegmark et al. 2001; Jaffe et al. 2000; Bond et al. 2000).  We begin by
constructing a four-dimensional grid in the parameter space using $\Omega_m$,
$\Omega_b$, $n_s$ the spectral index, and the Hubble constant.
We apply a weak prior with $H_0 = 69\pm{15}$km s$^{-1}$Mpc$^{-1}$, which is
consistent and more conservative than the final results of the Hubble Space
Telescope Key Project (Freedman et al. 2001). We then calculate the maximum
likelihood via ${\mathcal{L}} = e^{-\chi^2/2}$ where $\chi^2$ is calculated
using the fitting formulae given in Eisenstein \& Hu (1998).
In fitting these power spectra, 
we restrict ourselves to the range of $0.015\,<k\,<0.15$ using the uncorrelated
data of Figure 2.

The next issue is to find the global maximum in this multi--dimensional
likelihood space. Unfortunately, it is highly possible that this global
maximum does not lie on one of our grid points, so to guard against this, we
use the simplex method (see Press et al. 1992). We then marginalize over each
parameter separately by fixing it to the grid and allowing the other three
parameters to vary until we find the corresponding maximum. Our
methodology is very similar to that of Tegmark \& Zaldarriaga (2000) except
for our use of the simplex method to find the maximum likelihood.  Tegmark \&
Zaldarriaga fit a cubic spline to their likelihood grid to find the
maxima. However, this function can be ill--behaved if the surface is not smooth
({\it i.e.}  if
$\chi^2$ varies rapidly in the region of the minimum which is often the case).
Therefore, we advocate the use of the simplex method for future analyses
since, in principle, it is less dependent upon the actual likelihood surface.
We note that a proper marginalization requires integration over the likelihood
function. Tegmark et al. (2001) have shown that this is the same as the
maximization technique used herein if the likelihood functions are Gaussian,
which appears to be a reasonable approximation for our likelihood functions
(see Figure 3).

After marginalizing over the three power spectra separately, we combine the
likelihoods together to arrive at our final results. For each of the samples,
we allow the amplitude to be a free parameter. In this way, the bias parameter
does not explicitly enter into the calculation.  In Figure 3, we present the
Cash statistic for each marginalized parameter (Cash 1979): ${\rm Cash}_{i} =
-2ln\frac{{\mathcal{L}}_i}{{\mathcal{L}}_{max}}$, where ${\mathcal{L}}$ is the
maximum likelihood determined as a function of the fixed parameter $i$ and
allowing the other parameters to vary.  ${\mathcal{L}}_{max}$ is the global
maximum over all parameter space.  If the $log$ likelihood functions can be
well-approximated with a second order Taylor expansion, then the Cash
statistic becomes analogous to a $\chi^2$ distribution. Thus, when we
marginalize (i.e.  hold one parameter fixed allowing the others to vary), we
have one degree-of-freedom and our 95\% confidence limits are where our Cash
statistic crosses a value of $3.84$.  In Table 1, we present our final
estimates for the three cosmological parameters, $\Omega_o, \Omega_b,$ and $n$.
We also list similar recent results from latest CMB data (Tegmark
et al. 2001; Jaffe et al. 2000; Novosyadlyj et al. 2000). This table clearly
illustrates that our best fit values for these cosmological parameters are
fully consistent with these other analyses.

In Figure 4, we present the data in Figure 1 along with our best fit model
shown in Table 1. For comparison, we also show two favored zero baryon $n_s=1$
models with $\Gamma=0.25$ and $\Gamma=0.14$. Clearly, the baryon model is the
best representation given these three possibilities.

\begin{deluxetable}{ccccccc}
\tablenum{1}
\tablewidth{0pt}
\tablecaption{\bf Parameter Estimation Results}
\tablehead{
\colhead{$\Omega_mh^2$} & \colhead{$\Omega_bh^2$} & \colhead{$n_s$} & \colhead{Confidence} &
\colhead{Strong Priors} & \colhead{Data Used\tablenotemark{a}} & \colhead{Reference}}
\startdata
$0.12^{+0.03}_{-0.02}$  & $0.029^{+0.011}_{-0.015}$ & $1.08^{+0.17}_{-0.20}$ & $2\sigma$ & & LSS & This Work \nl
$0.22^{+0.07}_{-0.06}$  & $0.030^{+0.004}_{-0.004}$ & $0.99^{+0.07}_{-0.06}$ & $1\sigma$ & $\Omega_{tot} = 1$ &  CMB & Jaffe et al. (2000) \nl
$0.16^{+0.02}_{-0.01}$  & $0.032^{+0.004}_{-0.004}$ & $1.00^{+0.09}_{-0.06}$ & $1\sigma$ & & CMB + LSS & Jaffe et al. (2000) \nl
$0.15^{+}_{-0.15}$  & $0.054^{+0.05}_{-0.03}$ & $1.43^{+}_{-0.52}$ & $2\sigma$ & & CMB & Tegmark et al. (2001) \nl
$0.23^{+0.16}_{-0.12}$  & $0.028^{+0.008}_{-0.009}$ & $0.96^{+0.20}_{-0.10}$ & $2\sigma$ & &  CMB +LSS & Tegmark et al. (2001) \nl
$0.15^{+0.04}_{-0.02}$  & $0.02$ & $0.92^{+0.08}_{-0.09}$ & $2\sigma$ & $\Omega_b$ fixed &  CMB +LSS & Tegmark et al. (2001) \nl
$0.20^{+0.29}_{-0.11}$  & $0.019^{+0.04}_{-0.008}$ & $1.12^{+0.27}_{-0.30}$ & $1\sigma$ & $\Omega_{tot}$ = 1 &  CMB + LSS + $\sigma_8$ & Novosyadlyj et al. (2000) \nl
 & & & & & $Ly-\alpha$ forest + bulk flows & \nl

\enddata
\end{deluxetable}

\section{Conclusions}

We present in this paper new evidence for the detection of statistically
significant fluctuations in the matter power spectrum. The most natural
explanation for these fluctuations are baryonic oscillations in a Cold Dark
Matter universe as outlined in Eisenstein et al. (1998) and Eisenstein \& Hu
(1998). Using this cosmological model, we have measured $\Omega_m$,
$\Omega_b$, and $n_s$, finding values that are fully consistent with those
presently favored by the recent CMB experiments (see Table 1).  This agreement
is primarily due to the extra power seen on large scales (small $k$) as
well as the features detected in all three power spectra.
In the near future, surveys like the SDSS and
2dF Galaxy Redshift Survey (2dFGRS; Colless et al. 2000) will allow for a more
detailed analysis of these baryonic features as well as providing more
powerful constraints on the cosmological parameters than those presented
here (e.g. Percival et al. 2001). 
However, we do note that on large scales, the volume sampled by the
Abell/ACO catalog discussed herein will remain unrivaled even after the main
SDSS and 2dFGRS galaxy redshift surveys are completed and will thus remain,
for some time, an important database for studying the large--scale structure
in the Universe. However, the SDSS Bright Red Galaxy (BRG; York et al. 2000)
redshift survey will supercede all these surveys in terms of volume since it
will provide a pseudo--volume--limited sample of galaxies out to $z\simeq0.45$
carefully selected to sample the power spectrum of mass over as large a range
of scales as possible.

{\noindent {\bf Acknowledgments} ~
We are indebted to Chris Genovese and Larry Wasserman for their help with
FDR. We thank Christopher Cantaloupo, Adrian Melott, Wayne Hu,
Peter Coles, Alex Szalay, Daniel
Eisenstein, Andrew Jaffe and John Beacom for helpful advise and suggestions throughout this
work.}

\begin{figure}
\plotone{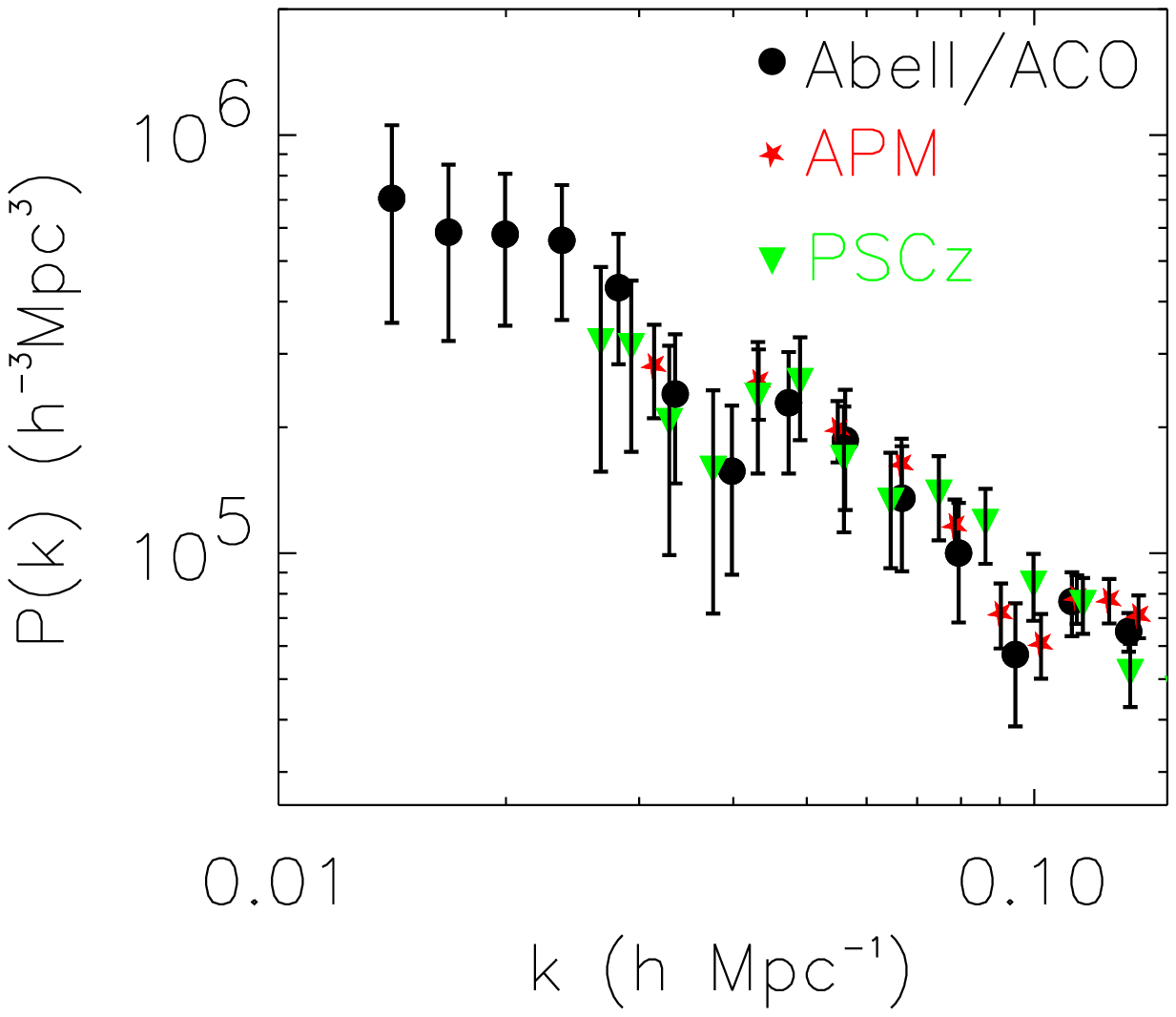}
\caption[]{The power spectra for the three samples utilized in this work.
The triangles are the PSCz galaxies, the stars are the APM groups
and clusters, and the circles are the Abell/ACO rich clusters.
The bottom panel shows $P(k)$ after a constant relative bias is applied
to the data sets. The shift corresponds to
$b = 1.5$ and $b = 3.6$ for
Abell/ACO versus APM and Abell/ACO versus the PSCz respectively.
}
\end{figure}

\begin{figure}
\plotone{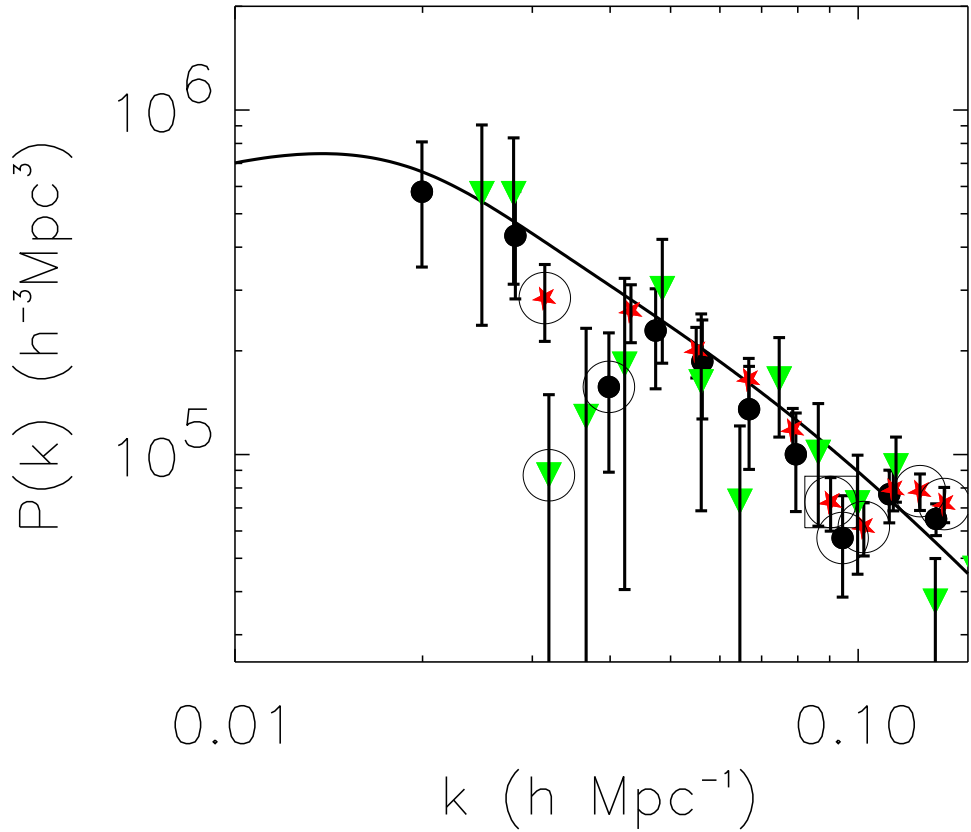}
\caption[]{Here, we show the amplitude shifted power spectra
for the three samples of uncorrelated data. The points highlighted with a circle
denote rejections with $\alpha = 0.25$ (e.g. a quarter of
the rejections may be mistakes). The points highlighted by squares
are for $\alpha = 0.10$ (e.g.  a tenth of the rejections
may be mistakes). The analysis utilizes our best-fit model
with the baryon wiggles removed as the null hypothesis.
By controlling the false discovery rate, we can say with statistical confidence
that the two ``valleys'' are detected as features in the power spectra.}
\end{figure}

\begin{figure}
\plotone{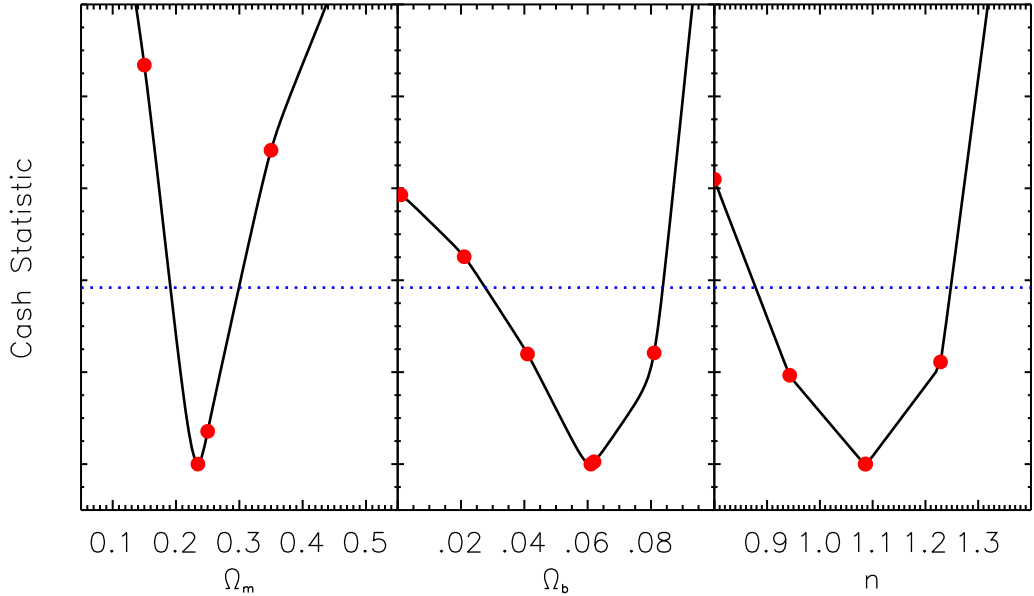}
\caption[]{The Cash statistic for the marginalized parameter estimations.  The
line corresponds to a 95\% confidence region.}
\end{figure}

\begin{figure}
\plotone{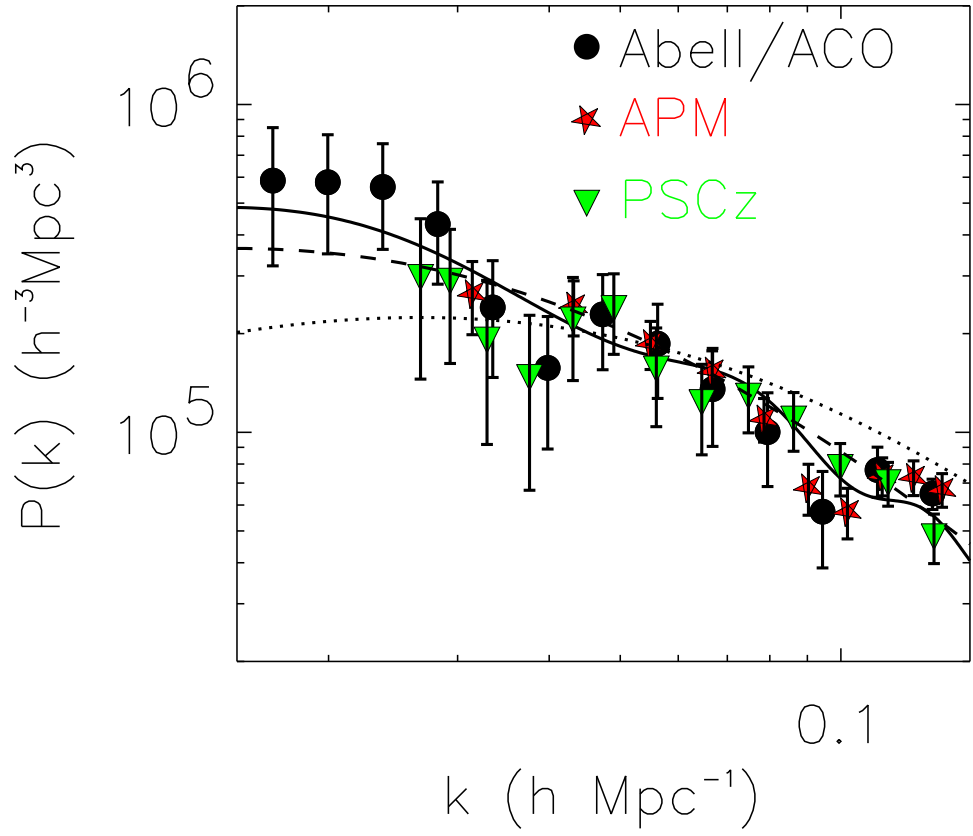}
\caption[]{The solid line is the best fit model ($\Omega_m = 0.24$, $\Omega_b
= 0.061$, and $n_s = 1.08$ with $H_0 = 69$; see Table 1) plotted with the data
from Figure 1 (bottom).
The dotted and dashed lines are zero
baryon $n_s=1$ models with $\Gamma=0.14$ and $\Gamma=0.25$ respectively.}
\end{figure}

\end{document}